\RequirePackage{fix-cm}
\documentclass[twocolumn,epjc3]{svjour3}  
\smartqed  
\RequirePackage{graphicx}
\RequirePackage[version=4]{mhchem}
\RequirePackage{siunitx}
\RequirePackage{amssymb}
\RequirePackage{mathptmx}      
\usepackage{wasysym}
\usepackage[symbol]{footmisc}
\usepackage{anyfontsize}
\usepackage{booktabs}
\usepackage{caption} 
\usepackage{amsmath}
\RequirePackage{hyperref}
\hypersetup{
  breaklinks=true,
  colorlinks=true,
  urlcolor=blue
}
\journalname{Eur. Phys. J. C}
\begin{document}

\title{First use of large area SiPM matrices coupled with NaI(Tl) scintillating crystal for low energy dark matter search}

\titlerunning{The ASTAROTH Experiment}
\author{Edoardo Martinenghi\thanksref{e1,addr1}
\and Valerio Toso\thanksref{addr1,addr2}
\and Fabrizio Bruno Armani\thanksref{addr1,addr2}
\and Andrea Castoldi\thanksref{addr1,addr3}
\and Giuseppe Di Carlo\thanksref{addr4}
\and Luca Frontini\thanksref{addr1}
\and Niccol\`o Gallice\thanksref{addr1,addr2,addr5}
\and Chiara Guazzoni\thanksref{addr1,addr3}
\and Valentino Liberali\thanksref{addr1,addr2}
\and Alberto Stabile\thanksref{addr1,addr2}
\and Valeria Trabattoni\thanksref{addr1,addr2}
\and Andrea Zani\thanksref{addr1}
\and Davide D'Angelo\thanksref{e2,addr1,addr2} 
}                     
\thankstext{e1}{e-mail: edoardo.martinenghi@mi.infn.it}
\thankstext{e2}{e-mail: davide.dangelo@mi.infn.it}
\institute{INFN - Sezione di Milano, via Celoria 16, 20133 Milano, Italy\label{addr1}
\and Dipartimento di Fisica, Universit\`a degli Studi di Milano, via Celoria 16, 20133 Milano, Italy\label{addr2}
\and Dipartimento di Elettronica, Informazione e Bioingegneria (DEIB), Politecnico di Milano, piazza Leonardo da Vinci 32, 20133 Milano, Italy\label{addr3}
\and INFN - Laboratori Nazionali del Gran Sasso (LNGS), via G. Acitelli 22, 67100 Assergi, Italy\label{addr4}
\and Brookhaven National Laboratory, PO 5000, Upton, NY 11973, USA\label{addr5}}
\date{Received: 30 July 2025 / Accepted: 8 December 2025}
%
\maketitle
\abstract{
The long-standing claim of dark matter detection by the DAMA experiment remains a crucial open question in astroparticle physics. A key step towards its independent verification is the development of NaI(Tl)-based detectors with improved sensitivity at low energies. 
The majority of NaI(Tl)-based experiments rely on conventional photomultiplier tubes (PMTs) as single photon detectors, which present technological limitations in terms of light collection, intrinsic radioactivity and high noise contribution at keV energies.
ASTAROTH is an R\&D project developing a NaI(Tl)-based detector where the scintillation light is read out by silicon photomultipliers (SiPMs) matrices. SiPMs exhibit high photon detection efficiency, negligible radioactivity, and, most importantly, a dark noise nearly two orders of magnitude lower than PMTs, when operated at cryogenic temperature. To this end, ASTAROTH features a custom-designed cryostat based on a bath of cryogenic fluid, able to safely operate the detector and the read-out electronics down to about 80~K. 
This work reports the first experimental characterization of an approximately 360~g NaI(Tl) detector read out by a large area (\qtyproduct{5 x 5}{\centi\metre}) SiPM matrix. The net photoelectron yield obtained with a preliminary configuration is approximately 4.5 photoelectrons/keV after crosstalk correction, which is rather promising in light of several planned developments.
The signal-to-noise ratio and the energy threshold attainable with SiPMs is expected to improve the sensitivity for dark matter searches beyond the reach of current-generation PMT-based detectors. 
This result is the first proof of the viability of this technology and sets a milestone toward the design of future large-scale experiments.}


%
\maketitle

\section{Introduction}
\label{sec:intro}

Among the open questions in modern physics, the nature of dark matter remains unknown. Astronomy offers several observations that could be explained with the introduction of dark matter: galaxy rotation curves~\cite{Sofue2001}, gravitational lensing~\cite{Clowe2006} and the large-scale structure of the cosmos~\cite{Freese2009,Boylan2009}. 
The presence of a yet-undetected mass, possibly comprised of unknown particles, emerges as the most plausible explanation of such phenomena~\cite{Bertone2010}.

In recent years, a variety of dark matter candidates have been proposed, including weakly interacting massive particles (WIMPs)~\cite{Goodman1985}, 
which are currently the focus of large research efforts in the field \cite{cirelli2024}.
Several experiments have sought to detect the presence of WIMPs by observing recoil energy from target nuclei.
The recoil energy is expected to be of the order of a few kiloelectronvolts, and interactions are extremely rare ~\cite{Lewin1996}. 
Therefore, it is crucial to develop a detector with an ultra-low background and low energy threshold, which represents the greatest challenge in direct detection experiments.
To date, the sole positive claim of dark matter interaction, reported over the last 20 years, is attributed to the DAMA experiment, which observed an annual modulation of the event rate from an array of NaI(Tl) scintillating crystals located underground in Italy at the Laboratori Nazionali del Gran Sasso (LNGS)~\cite{Freese2013}, with events detected in the recoil energy range of \qtyrange[range-units=single,range-phrase=-]{1}{6}{keV_{ee}} \cite{Bernabei2013,Bernabei2018,Bernabei2021}.

This result cannot be easily reconciled with the negative findings of several more sensitive direct detection experiments (e.g., XENON~\cite{Xenonnt}, PandaX~\cite{PandaX2025}, LZ~\cite{LZ2024}, Dark Side \cite{Agnes2018_2}) within the standard WIMP interaction model. 
However, such detectors use different target materials and techniques, and comparisons require several assumptions. For example, a possible dependence of WIMP interactions on specific target nuclei has been proposed \cite{Catena2016}. 
An independent verification with the same target material and technique is essential\footnote{Request emphasized by the Astroparticle Physics European Consortium (APPEC) in its 2017-2026 roadmap~\cite{appec}.}.

Considering this aspect, few NaI(Tl)-based experiments have been conducted, none of which have yet provided a definitive result due to limited exposure and statistical sensitivity (e.g., ANAIS \cite{anais2025}, COSINE~\cite{cosine2025}).  Other projects are also in preparation (e.g., SABRE \cite{Antonello2019,Calaprice2022}, PICOLON \cite{Fushimi2022}).

All of the above detectors share a common design, where scintillation light is read by photomultiplier tubes (PMTs).
Although PMTs represents a well established technology in the field, there are significant limitations in their application with NaI(Tl) for dark matter direct detection. Their instrumental noise rates are orders of magnitude higher than the rate of scintillation events at the keV energies of interest for dark matter. Although pulse shape analysis techniques can help discriminate these two categories of events, they struggle to achieve a reasonable efficiency below 1~keV \cite{Mariani21}. 

The suppression of PMT-related noise from the region of interest, namely up to \qty{6}{\kilo\electronvolt}, would significantly enhance the signal-to-noise ratio and allow for the exploration of the sub-keV region. Moreover, PMTs are relatively large and their constituent materials intrinsically non-radiopure. Even low radioactivity models cannot achieve less than \SI{5}{\milli\becquerel} \cite{PandaXPMTS2024}. To limit the impact of this background, NaI(Tl) crystals of considerable mass are required, but manufacturers struggle to reproducibly obtain products that are both several kilograms in mass and ultra-highly radio-pure. 

Despite being a relatively recent technology, silicon photomultipliers (SiPMs) \cite{Dinu2016} have undergone impressive developments in the last few years~\cite{Buzhan2003}. They are ideal candidates  to overcome the limitations of PMT readout of NaI(Tl) scintillation light at low energies. SiPMs offer several advantages, of which three are particularly relevant to this application:
(i) primary dark noise in SiPMs, arising from spontaneous generation of electron–hole pairs in the semiconductor, decreases rapidly with temperature and can be up to two orders of magnitude lower than that of PMTs when operated down to \qty{80}{\kelvin}~\cite{Biroth2015}. 
On the other hand, only a few PMT models are functional at low temperature and do not exhibit a significant reduction in dark noise when cooled.
(ii) SiPMs, composed predominantly of silicon and fabricated using precision microelectronic processes, are very thin, compact, and can be intrinsically radio-pure devices. 
(iii) SiPMs have comparatively high photon detection efficiency (PDE); for example near-ultraviolet (NUV) devices show a PDE around \qty{65}{\percent} at \qty{420}{\nano\metre}~\cite{Merzi_2023}, which is the peak emission of NaI(Tl) crystals \cite{Sibczynski2011}. 
For comparison, PMTs PDE at this wavelength typically settles around \qtyrange[range-units = single,range-phrase=-]{30}{35}{\percent} (e.g. Hamamatsu R11065-20 \cite{SABRESouthPMTS2025}). 

With the ASTAROTH project \cite{Dangelo2023}, the goal is to
demonstrate the feasibility of coupling SiPM matrices with cubic NaI(Tl) crystals for direct dark matter detection, employing an innovative approach targeting the keV and sub-keV energy ranges. To fully exploit the performance of SiPMs, the detector must be operated in a cryogenic setup.
In this paper, the prototype design of the ASTAROTH project,
located at the Laboratorio di Acceleratori e Superconduttività Applicata (LASA) in Segrate, Italy, is presented.
Initial results are also presented, representing —to the best of current knowledge— the first cryogenic operation of an approximately 360 g NaI(Tl) detector read out by a large-area (\qtyproduct{5 x 5}{\centi\metre}) SiPM matrix\footnote{The ANAIS+ project also pursues the use of SiPMs with NaI(Tl) detectors, although with a different technological strategy \cite{Anais_plus2024}, demonstrating the interest of the community.}. This sets a remarkable step toward the exploration of the sub-keV region of the energy spectrum, where new information remains to be uncovered \cite{Freese2013}.

In Section~\ref{sec:detector}, the NaI(Tl) detector design is introduced; Section~\ref{sec:cryo} describes the innovative cryostat developed and Section~\ref{sec:elec} details the cryogenic front-end readout of the SiPM matrix. Sections~\ref{sec:calib} and~\ref{sec:scint} present the detector characterization using laser and scintillation light, respectively, before the conclusions in Section~\ref{sec:conclusions}.

\section{Detector}
\label{sec:detector}

\begin{figure*}
\resizebox{\textwidth}{!}{%
\includegraphics{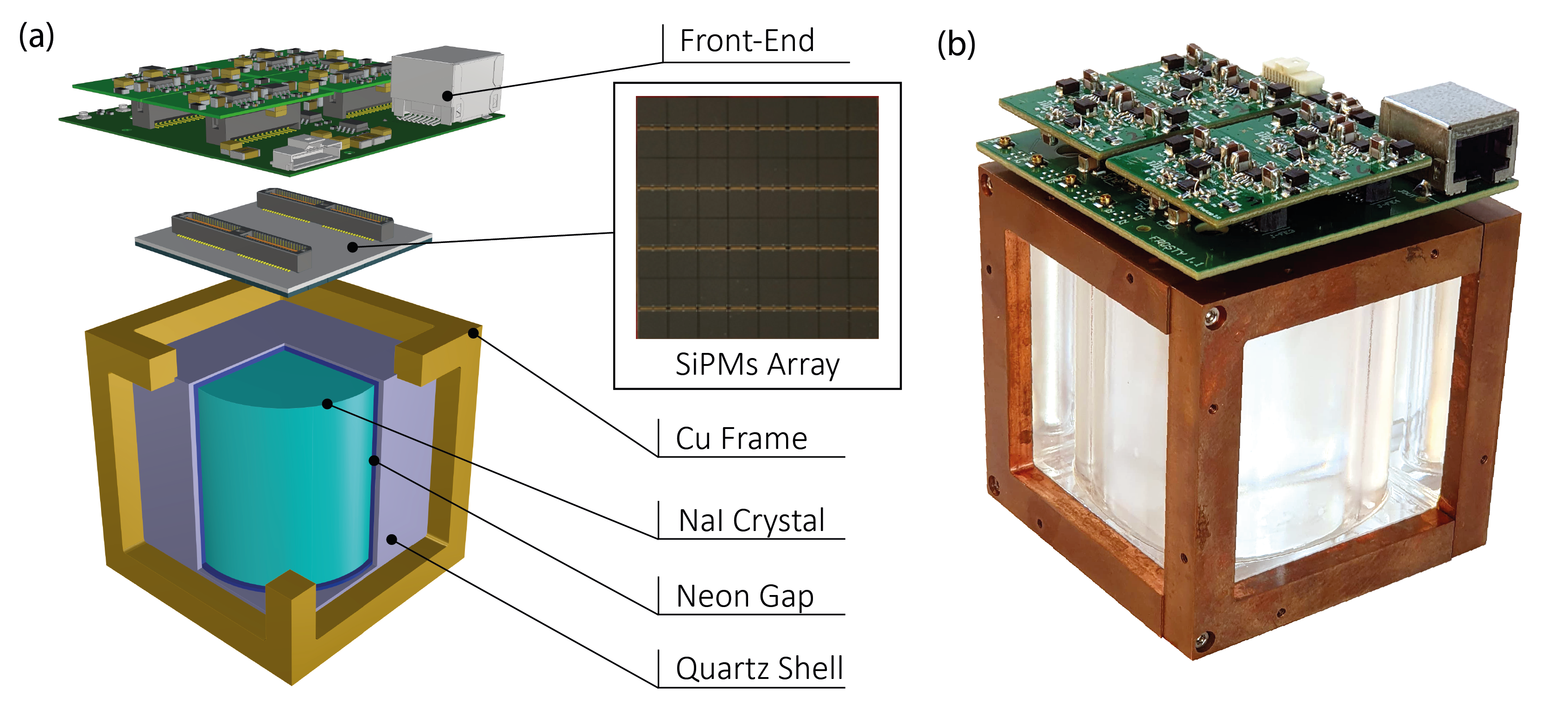}
}
\caption{(a) 3D design of the detector, showing how the NaI(Tl) crystal is encapsulated into the synthetic quartz shell. The copper frame allows coupling the SiPM matrix and the front-end electronics to the crystal. (b) Photo of the detector.}
\label{fig:detector}
\end{figure*}

The detector for the data presented in this article was assembled using a NaI(Tl) crystal within a fused silica case and a SiPM matrix covering one of the faces. Figure~\ref{fig:detector} presents a schematic view and photograph of the detection unit.

\subsection{Crystal}
\label{sec:crystal}

The crystal used is a thallium-doped NaI crystal with a standard dopant concentration of 750~ppm.  The emission spectrum peaks at \SI{420}{\nano\metre}, matching the sensitivity of NUV-class SiPMs. 
The light yield is traditionally believed to be close to \qty{40}{photons\per \kilo\electronvolt} at room temperature \cite{Knoll2010}, with a primary de-excitation time around \qty{250}{\nano\second}. \footnote{Some more recent studies report a significantly higher yield \cite{Sasaki2006}.}
The cryogenic temperature behavior of NaI(Tl) crystals is not fully understood yet, but some studies report a much slower response, with time constants of the order of \qty{1.5}{\micro\second} \cite{Sibczynski2011,Gallice23}.

The crystal, fabricated by Hilger Crystals (UK), consists of a cylinder
with both height and diameter of 50 mm. As NaI is highly hygroscopic, an airtight fused silica (synthetic quartz) shell was used, which is fully transparent at \SI{420}{\nano\metre} at room and cryogenic temperature. The shell was constructed by boring a cylindrical hole in a silica cube and bonding two flat lids to both ends. 
The crystal was sealed inside the enclosure by the manufacturer under a dry neon atmosphere, leaving it surrounded by neon gas at 1~bar pressure and thus preventing damage from possible humidity contamination. A uniform 1~mm gap was left between the cylindrical crystal and the inner surfaces of the quartz cube (i.e., at the top and bottom bases as well as along the lateral surface) in order to accommodate the thermal expansion mismatch of the two materials across cryogenic cycles.

The case architecture was recommended by the producer for this initial phase, due to its ease of production and reduced risk of leakage. However, in this configuration a significant fraction of photons can be lost due to total internal reflection at the crystal-neon interface. To quantitatively estimate this effect, a Geant4-based Monte Carlo optical simulation was developed, which showed that only \qty{35}{\percent} of the light reaches the photo-detectors ~\cite{Galli2022}. \footnote{To avoid this signal loss, a new solution based on epoxy resin has been developed and is currently being tested.}

The outer dimensions of the case are \qtyproduct{58 x 58 x 58}{\milli\metre}. An oxygen-free high-conductivity (OFHC) copper frame surrounds the quartz shell, allowing the coupling with the SiPM matrix and the readout electronics. 
In the present setup, the SiPM matrix is held at millimeter distance from the quartz surface while an optical coupling that remains transparent at low temperature is under study.

\subsection{SiPMs}
\label{sec:sipm}

NUV-HD-cryo low-field SiPMs produced by Fondazione \break Bruno Kessler (FBK) and specifically designed for cryogenic operation, were chosen for use in the current system. 

SiPMs measure \qtyproduct{6 x 6}{\milli\metre}, a cell pitch of \SI{30}{\micro\metre}, and exhibit the following performance, when operated in liquid nitrogen (77~K) with a \SI{7}{\volt} excess-bias: (i) PDE of \qty{55}{\percent} at \SI{420}{\nano\metre}; (ii) dark count rate (DCR) below 10$^{-2}$~\si{cps\per\milli\metre\squared}\footnote[2]{cps = counts per second}; (iii) afterpulsing probability lower than 20~\%. A detailed characterization of this technology is widely available in the literature~\cite{Acerbi2017}.
The matrix was assembled by FBK according to a jointly developed design. 
64 SiPMs, arranged in an \numproduct{8 x 8} configuration, are mounted on a single FR4 printed circuit board (PCB)(Figure~\ref{fig:detector} (a)). Electric contacts are provided by conductive glue for the cathode and by wire bonding for the anode. The total active area is \qtyproduct{48 x 48}{\milli\metre}. The variation of the breakdown voltage over the 64 devices was measured by the manufacturer to be less then 0.5~\%, thus ensuring uniform performances over the whole active area of the matrix.

A \qtyrange[range-units = single]{200}{300}{\micro\metre} layer of epoxy resin safeguards against accidental damage during coupling with the crystal. On the opposite side of the PCB, two high-pitch 80-pin connectors enable coupling to the electronic front-end, which is detailed in Section~\ref{sec:elec}. 

\section{Cryogenic setup}
\label{sec:cryo}

\begin{figure*}
\resizebox{\textwidth}{!}{%
\includegraphics{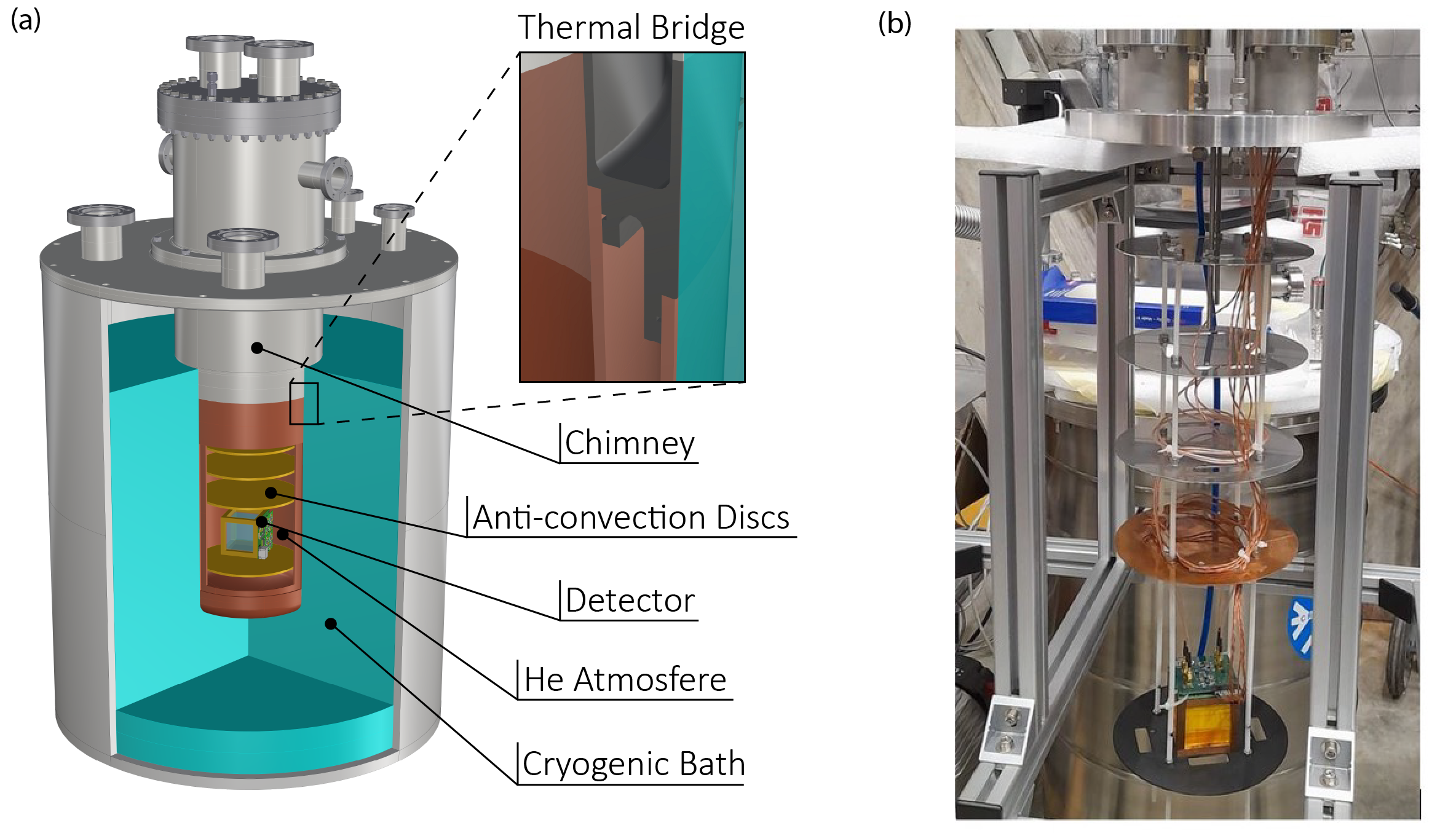}
}
\caption{(a) 3D design of the cryostat showing the cryogenic chamber, where the detector is hosted in a helium atmosphere (1~bar), surrounded by a cryogenic bath. A sequence of discs allows for the stratification of the inner gas and minimizes convection cycles along the height of the chimney, which is directly exposed to room temperature at its top. The inset details the section of the thermal bridge, responsible for cooling the inner volume. (b) Photograph of the inner crystal support structure, hanging from the main chimney flange and featuring the anti-convection disks, ready to be inserted into the cryogenic chamber.}
\label{fig:cryostat}
\end{figure*}

The innovative cryostat\footnote{The design of this cryostat is the result of an extensive study, where multiple simulations and cryogenic stress tests of the materials were performed by the Mechanical Design Services of INFN Milan and INFN LNGS, in collaboration with the authors~\cite{astaroth-simul}.} shown in Figure~\ref{fig:cryostat} was designed to operate SiPMs at low DCR conditions, at a tunable temperature from \qtyrange[range-units = single]{80}{150}{\kelvin} range. 
The upper limit is set at the point where the DCR of SiPMs is similar to that of a typical PMT (\qty{\sim 0.1}{cps\per\milli\metre\squared}) \cite{SABRESouthPMTS2025}, while the lower limit is set by the temperature of the cryogenic fluid employed. 
At \qty{80}{\kelvin} the DCR of the FBK SiPMs is already at the level of \qty[print-unity-mantissa = false]{\sim e-2}{cps\per\milli\metre\squared} and a lower temperature would bring no further advantage. 
The temperature dependence of the DCR of these SiPMs can be found in \cite{Acerbi2017}.

The cryostat exploits the cooling power from a cold fluid surrounding a cryogenic chamber that houses the detector. The chamber is made of a double-walled OFHC copper container, with an inner diameter of 214~mm, connected to a stainless steel (SS) chimney. OFHC copper is selected for its high thermal conductance and low radioactivity. 
The copper section of the chamber consists of two vacuum-insulated, 3~mm-thick walls, brazed to a stainless steel (SS) bridge element. The thermal bridge, which connects the Chimney to the copper part of the internal chamber, serves as the primary path for heat transport to the inner volume, since (1) thermal irradiation between the two walls has been calculated and found to be negligible, and (2) a series of thin SS disks along the chimney height promotes gas stratification, preventing large convection cycles. 

\begin{figure*}
\resizebox{\textwidth}{!}{%
\includegraphics{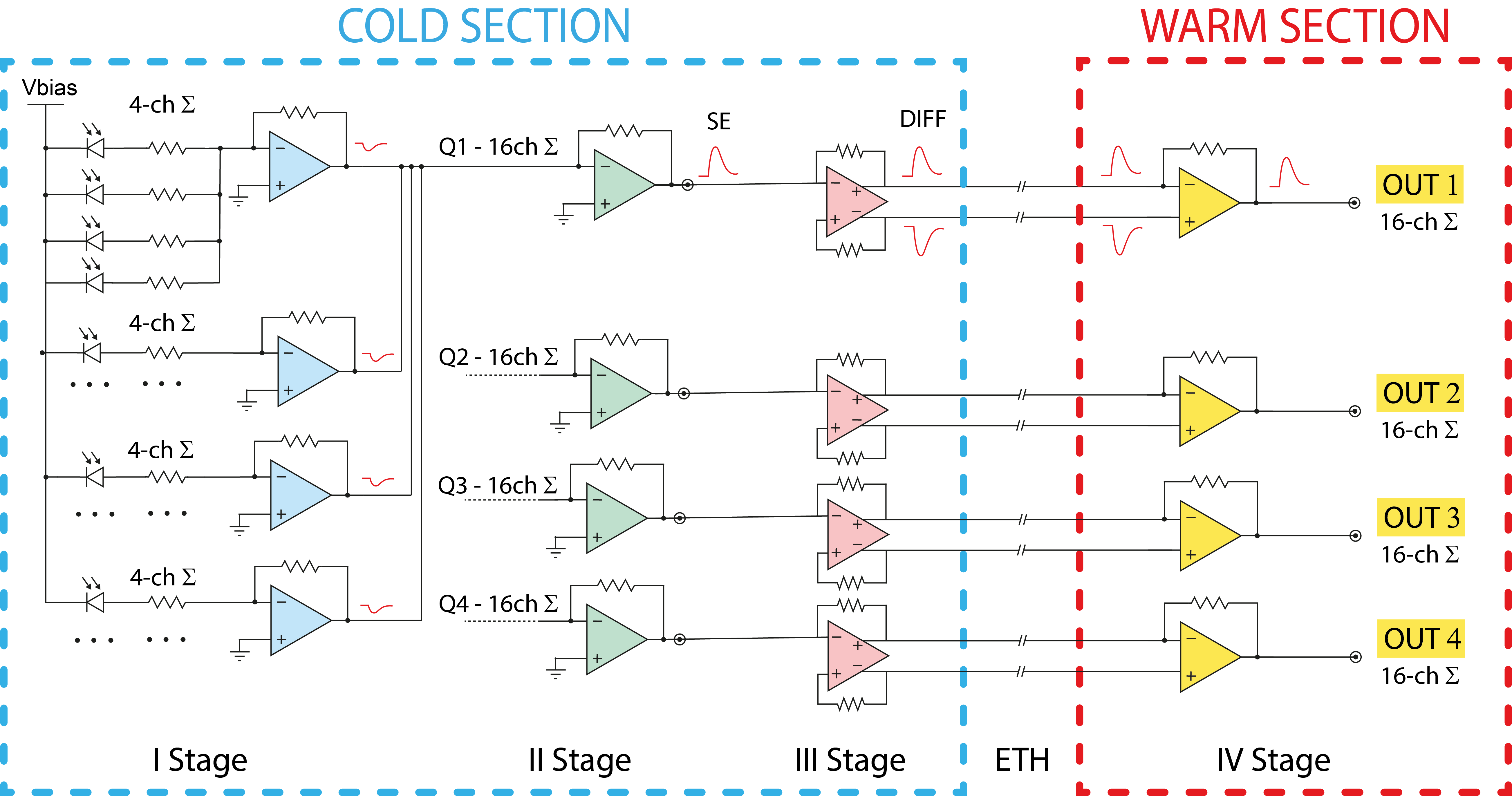}}
\caption{Block schematic of the electronics front-end. The \textit{cold section} sums the signals of the 64 SiPMs over two stages of amplification and converts them into four differential pairs, one for each quadrant of the matrix. Signals are then routed out of the cryostat on an Ethernet cable and converted back to single-ended by the \textit{warm section}; finally they are sent to data acquisition.}
\label{fig:front-end}
\end{figure*} 

The volume of the chamber and the chimney is filled with dry helium gas\footnote{99.999~\% purity level of helium gas employed.}, which guarantees the complete thermalization of the detector, prevents water condensation and provides the necessary thermal inertia. The cryostat is supplied by an external helium gas cylinder that, thanks to a pressure controller, maintains a constant pressure of approximately 1.1~bar absolute inside the chamber during cooling cycles.

The cool-down to \qty{80}{\kelvin} is deliberately slow (\SI{<20}{\kelvin\per\hour}) to avoid significant temporal or spatial temperature gradients within the crystal that could induce mechanical stress and eventually lead to cracks. 
After cool-down, the temperature can be slowly raised and stabilized by means of an internal 250~W heating element, driven by a temperature controller (Lakeshore Model 336 \cite{lakeshore}), to any working point within the target range, ensuring fluctuations lower than 0.1~K and no significant drift across data acquisition cycles of several hours. For this initial run, liquid nitrogen (LN$_2$) was used as the cooling fluid at 77~K due to its immediate availability. 

\section{Front-end Electronics} 
\label{sec:elec}
A custom front-end electronic system was developed to read out the signals from the SiPM matrices. The front-end is divided into a \textit{cold section} to be operated inside the cryogenic chamber, directly backing the SiPMs, and a \textit{warm section} that receives the signals outside the cryostat. The scheme of the front-end electronics is shown in Figure~\ref{fig:front-end}.

The design is based on commercial SiGe technology integrated circuits that were selected after several performance tests in LN$_2$. For passive components, metal film resistors and C0G radio frequency capacitors were used, which have been demonstrated to have stable characteristics over a wide temperature range (i.e. down to 4~K~\cite{Pan2005}). Moreover, thanks to the extensive availability of SiPM simulation models in the literature~\cite{Acerbi2019,Villa2015,Marano2014}, a comprehensive simulation of the circuit behavior was performed, which provided essential detail of the overall design. 
 
The \textit{cold section} is designed as a three-stage amplifier reading the 64 independent SiPMs of the matrix. The initial stage uses a transimpedance amplifier (TIA) configuration based on a LMH6626 \cite{lmh6626} operational amplifier (op-amp), which sums the signals from four SiPMs. 
The second stage sums four signals from the first stage, using a LMH6624 \cite{lmh6624} op-amp  in an inverting configuration. This leaves the SiPMs grouped into four separate quadrants (Q1, Q2, Q3 and Q4), each providing a signal corresponding to 16 devices. 

In order to route these analog signals along the chimney and towards the data acquisition (DAQ) system, located outside the cryostat, without picking up significant electromagnetic noise, a third stage performs a conversion of the signals into differential pairs, which is obtained by means of THS4541 \cite{ths4541} fully differential op-amps. A CAT7 ethernet cable, made of four individually shielded twisted pairs, brings the signals out to the \textit{warm section} placed in close proximity, approximately 1~m, to the cryostat. Here, differential signals are converted back to single-ended ones for acquisition and digital sampling.

\section{System assessment and calibration}
\label{sec:calib}

The characterization and calibration of the SiPM matrix and read-out electronics is reported in this section, assessing its response to laser pulses before coupling it to the crystal.

\begin{figure}
\resizebox{\columnwidth}{!}{%
\includegraphics{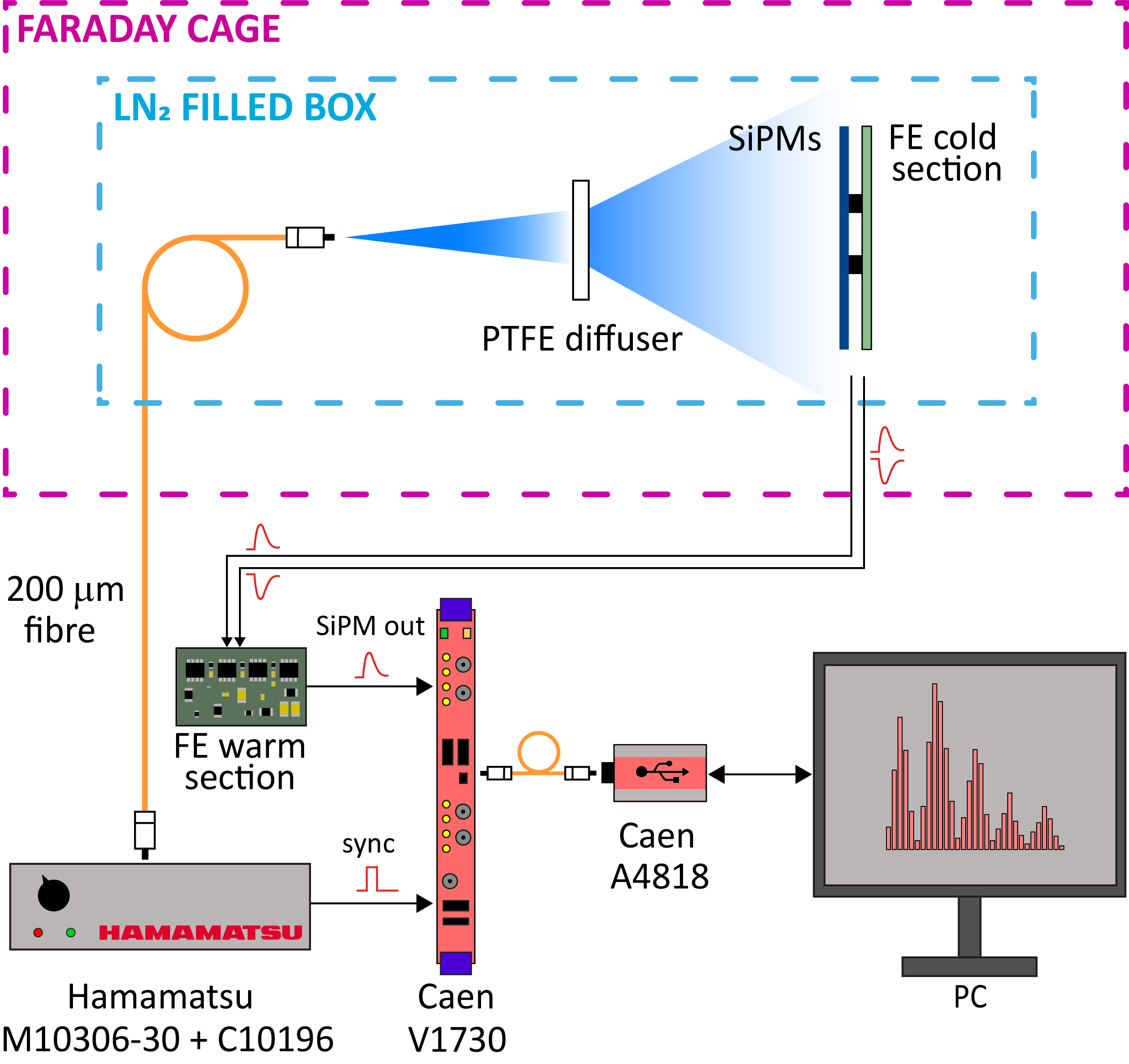}
}
\caption{Schematic diagram of the calibration setup; details are described in the text.}

\label{fig:laser}
\end{figure}

\subsection{Calibration setup}
\label{sec:calib_set-up}

Figure \ref{fig:laser} illustrates the setup employed for the calibration. To emulate the operating conditions expected in the cryostat, the SiPM matrix and the \textit{cold section} of the electronics front-end were immersed in a box of open-air evaporating \ce{LN2}, which maintained a stable temperature of \SI{77}{\kelvin}. To minimize electromagnetic interference and optical contamination from the environment, the box was placed inside a Faraday cage under completely dark conditions.

To generate precise optical pulses for the SiPM calibration, a picosecond-pulsed laser with a wavelength of 405~nm (Hamamatsu M10306-30 laser + C10196 controller \cite{hamamatsu}) was used as light source. The repetition rate was kept below 100~Hz, as signal-related noise in cryogenic SiPMs — such as afterpulsing — is typically exhausted within \qty{1}{\milli\second} after the initial event~\cite{Acerbi2017}. This slow repetition rate ensures that such noise remains negligible during acquisition. The laser intensity was adjusted to operate the SiPMs in the single-photon detection regime.

To guarantee uniform illumination, the laser optical output was connected to a multimode optical fiber with a \qty{200}{\micro\metre} core and numerical aperture of 0.39, whose other end was positioned inside the \ce{LN2} box at approximately 50~cm distance from the photodetector.
A polytetrafluoroethylene \break (PTFE) sheet was placed 10~cm in front of the SiPM as a light diffuser. The \textit{warm section} of the electronics front-end was instead positioned close to the data acquisition system, which employs an 8-channel, 14-bit, 500~MS/s digitizer (CAEN V1730) \cite{caenv1730}. Each quadrant of the matrix is read out by a dedicated channel. The board optical link (CAEN CONET2 protocol) is converted to USB~3.0 by a CAEN A4818 \cite{caena4818} interface card, providing a bandwidth of 80~MB/s to a connected personal computer (PC). The acquisition of the SiPM signal output was then triggered by the synchronization signal of the laser.

\subsection{Signal output}
\label{sec:calib_signal}
The output signal from a SiPM is proportional to the number of firing microcells; therefore, when multiple photons are simultaneously absorbed in different microcells, the device signal increases proportionally. 
Similarly, combining and summing the signals from multiple SiPMs results in a signal amplitude that is directly proportional to the total number of photons detected over the whole surface.

\begin{figure}
\resizebox{\columnwidth}{!}{%
\includegraphics{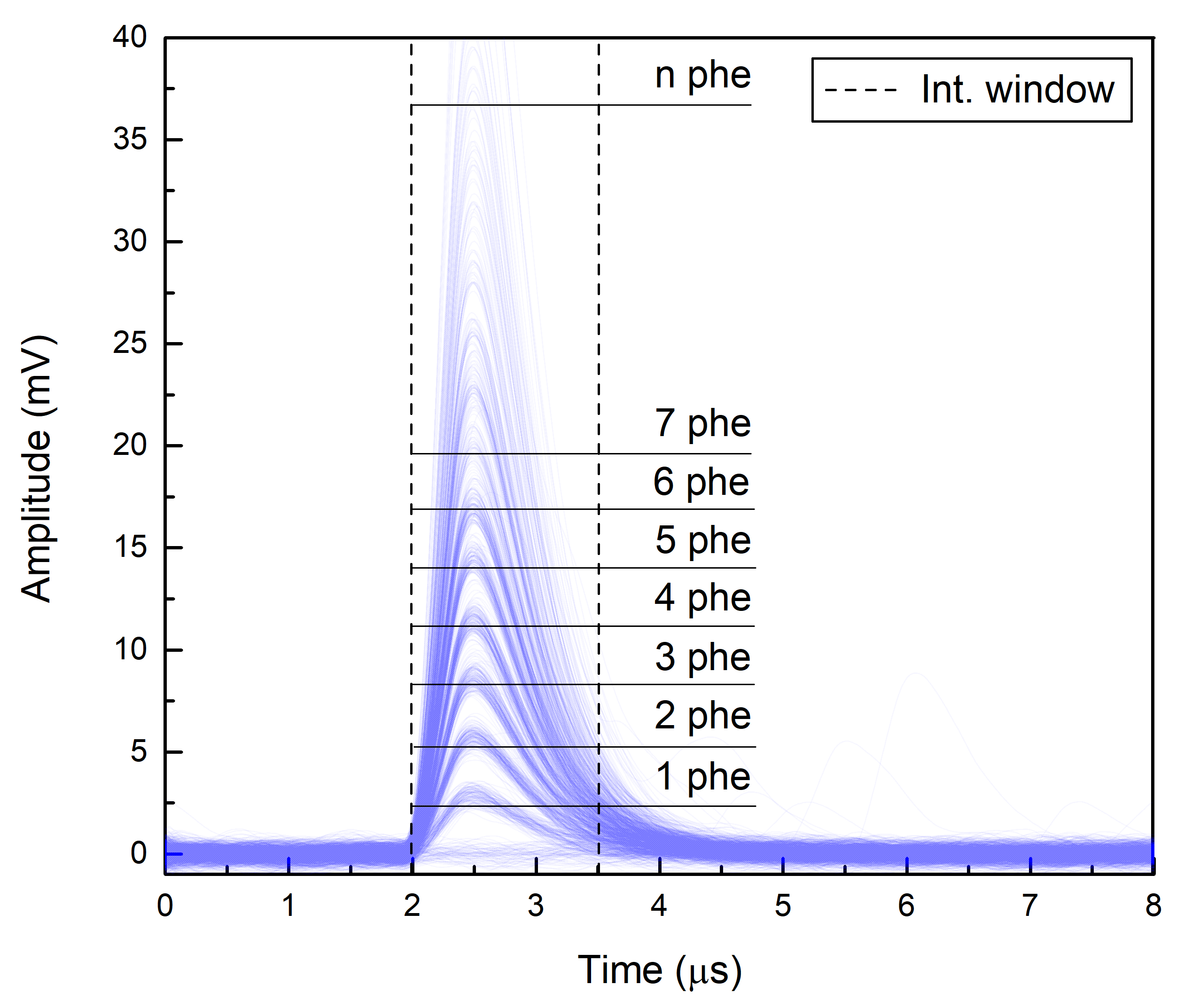}
}
\caption{\num{1000} baseline-subtracted waveforms acquired while illuminating the SiPMs with a low intensity laser. The pulse amplitude linearly corresponds to an increasing number of detected photoelectrons (phe). The dashed lines show the charge integration window, which was selected to integrate at least the 95~\% of the charge as described in the text.}
\label{fig:persistency}
\end{figure}

Figure~\ref{fig:persistency} shows \num{1000} waveforms acquired on one quadrant while illuminating the SiPMs with low intensity laser light \footnote{Nearly identical plots relative to the other three quadrants are not shown.}.
The waveform baseline was estimated as the mode over the \qty{2}{\micro\second} preceding the photon pulse and subtracted individually for each waveform. 

In the current configuration, a single-photon event produces a pulse with an amplitude of approximately \qty{2.5}{\milli\volt}, while in multi-photon events, the amplitude increases linearly with the photon count. The shape of the signal primarily depends on the SiPM parasitic capacitance and the front-end bandwidth, resulting in a rise time of approximately \qty{0.2}{\micro\second} an exponential decay time of about \qty{0.5}{\micro\second}.

\begin{figure}
\resizebox{\columnwidth}{!}{%
\includegraphics{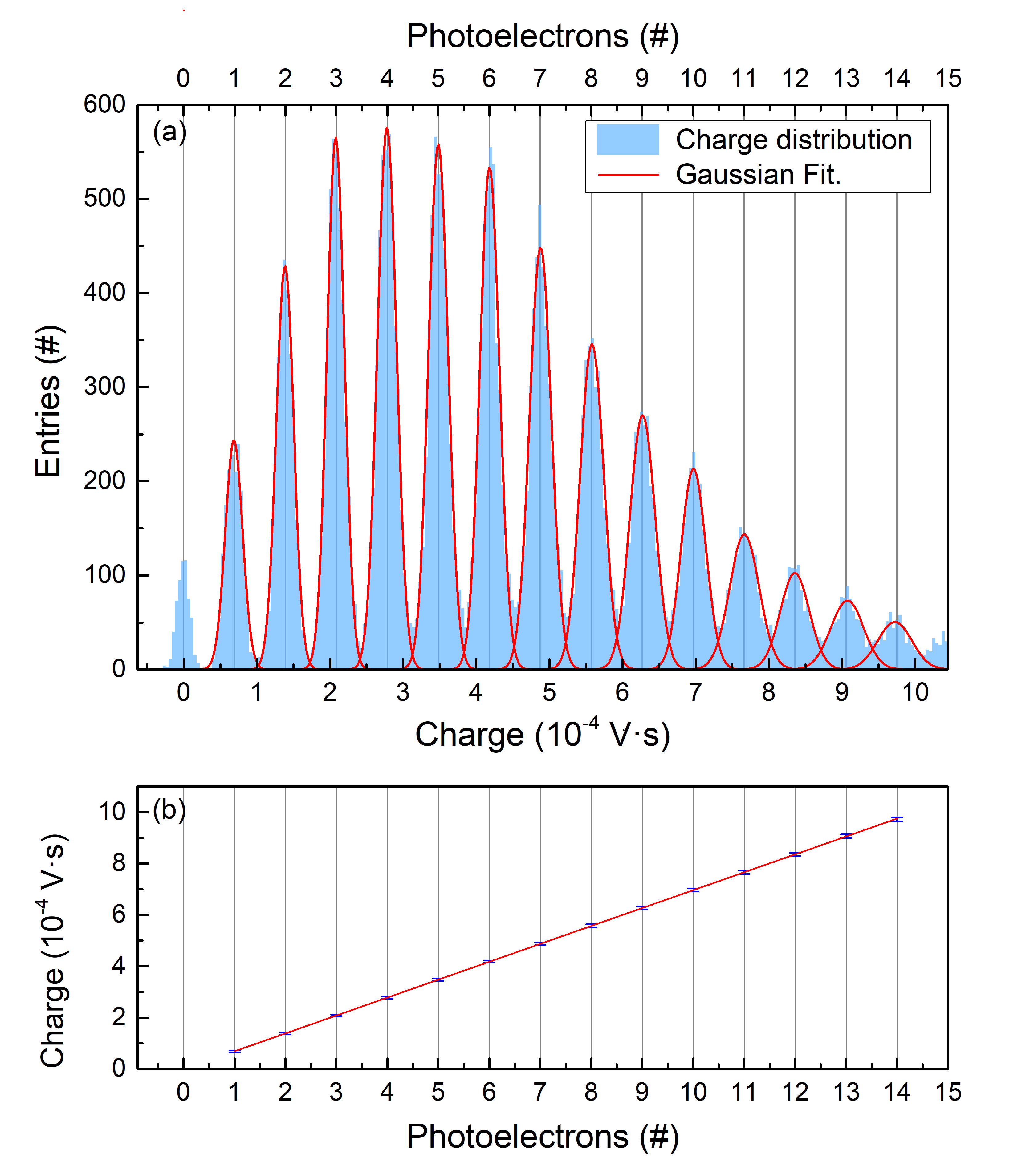}
}
\caption{(a) Charge histogram of \num{40000} waveforms acquired with low intensity laser light on one quadrant. Each photoelectron peak is fitted with a Gaussian profile. The upper x-axis is computed using the calibration parameter of the channel. (b) Peak positions: the linear fit shows an excellent linearity and allows to compute the calibration parameter of the channel.}
\label{fig:calib}
\end{figure}

\begin{figure*}
\resizebox{\textwidth}{!}{%
\includegraphics{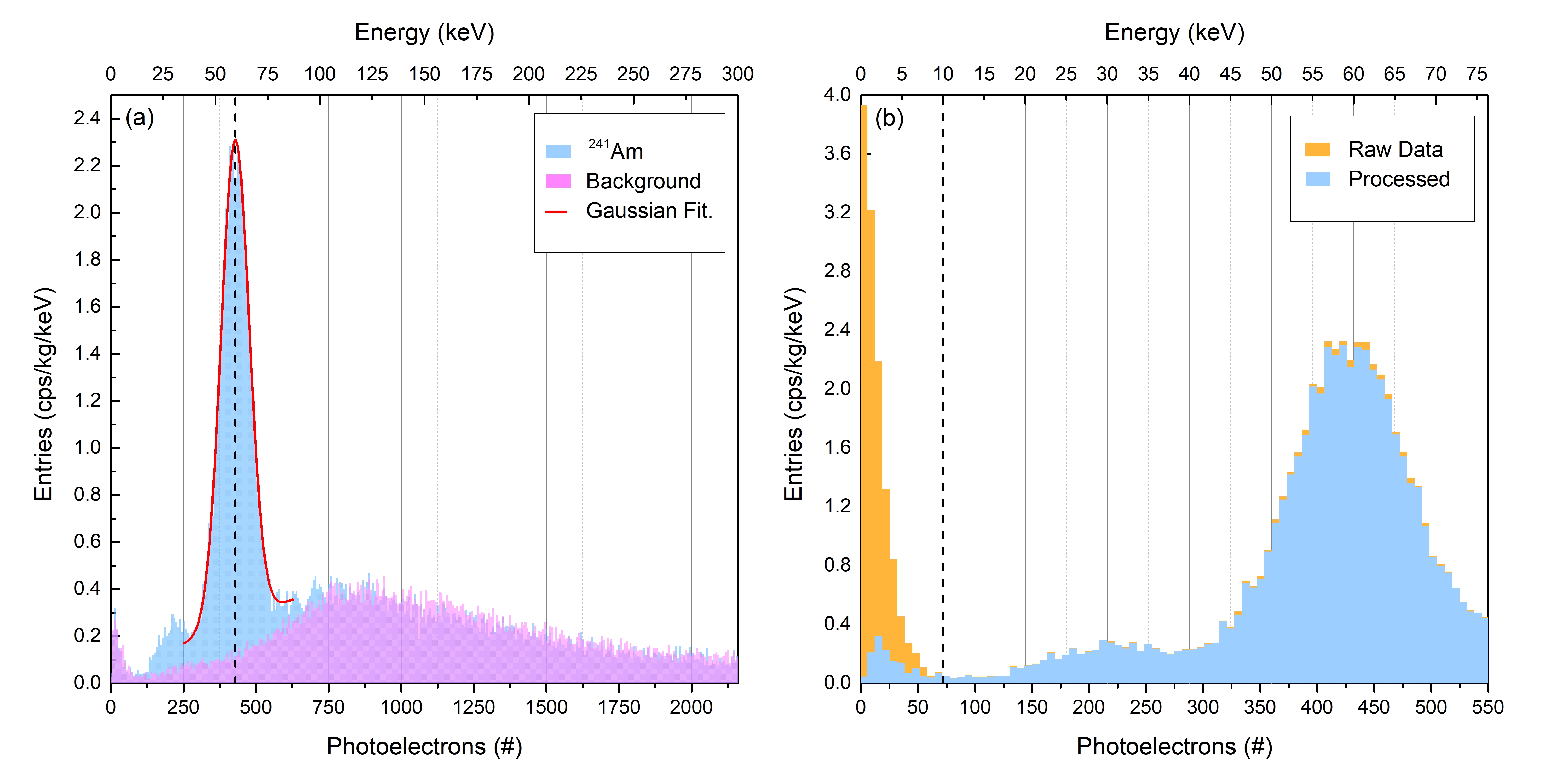}
}
\caption{(a) Energy spectra with (blue) and without (magenta) the \textsuperscript{241}Am source. The main peak is due to the 59.5~keV gamma and is fitted with a gaussian, obtaining a gross photoelectron yield of 7.2~phe/keV (4.5~phe/keV after crosstalk correction), which is then used to compute the upper x-axis scale. A secondary structure, due to x-ray emissions from \textsuperscript{237}Np (a product of \textsuperscript{241}Am decay chain), is visible around 220~phe / 30~keV. (b) Energy spectrum acquired with the \textsuperscript{241}Am source before (yellow) and after (blue) the data quality selection described in the text.}
\label{fig:source}
\end{figure*}

\subsection{Charge calibration and linearity}
\label{sec:calib_calib}

Each quadrant of the system measuring the charge collected per photoelectron was calibrated computing the integral of the acquired waveforms. Since the signal has a time constant of approximately \qty{0.5}{\micro\second}, the integration was performed over a \SI{1.5}{\micro\second} window starting from the onset of the pulse (dashed lines in Figure 4). The choice of the integration window is aimed at achieving an almost complete integration of the signal, which is crucial for obtaining a robust signal-to-noise ratio (SNR) with respect to the electronic noise of the system. On the other hand, using a larger integration window would result in over-integration of the electronic noise, leading to a deterioration of the SNR. By checking how charge varied when expanding the window up to \SI{3.0}{\micro\second}, it was estimated that over 95~\% of the pulse was exhausted within the chosen duration.

Figure~\ref{fig:calib} (a) shows the charge histogram for one of the quadrants, obtained by integrating \num{40000} waveforms\footnote{Nearly identical histograms for the other three quadrants are not shown.}.
The first peak, centered at null charge, is the \emph{pedestal} corresponding to the electronic noise of the channel. 
Subsequent peaks correspond to an increasing photon count.
An algorithm to identify the peaks and fit them individually with Gaussian profiles was developed.
Centroids of the profiles were used to construct Figure~\ref{fig:calib} (b). 
A linear fit was performed to compute the charge calibration parameter which in turn was used to build the upper x-axis in Figure~\ref{fig:calib} (a).
The fit in Figure~\ref{fig:calib} (b) also shows the excellent linearity of the device, with an adjusted R-Square of 0.99999.  

The SNR of each channel is computed as $(\mu_1 -\mu_0)/\sigma_0$, where $\mu_0$ and $\mu_1$ are the positions of the noise \emph{pedestal} and of the single-photon response peak, respectively, and $\sigma_0$ is the RMS of the \emph{pedestal}.  
The charge calibration parameters and SNRs for the four quadrants are summarized in Table~\ref{tab:charge_calib}:

\begin{table}[h!]
\centering
\begin{tabular}{|c|c|c|}
\toprule
\textbf{Quadrant} & {\textbf{Charge calib. (Vs/phe)}} & {\textbf{SNR}} \\
\midrule
1 & $(6.88~\pm~0.01)~\cdot~10^{-5}$ & $4.5~\pm~0.2$ \\
2 & $(7.47~\pm~0.01)~\cdot~10^{-5}$ & $4.5~\pm~0.2$ \\
3 & $(6.52~\pm~0.01)~\cdot~10^{-5}$ & $5.1~\pm~0.2$ \\
4 & $(6.97~\pm~0.01)~\cdot~10^{-5}$ & $4.5~\pm~0.2$ \\
\bottomrule
\end{tabular}
\caption{Charge calibration and SNR for each quadrant.}
\label{tab:charge_calib}
\end{table}

When dealing with SiPMs the occurrence of crosstalk \cite{Acerbi2017} can lead to an overestimation of the actual number of detected photons. To quantify this effect, the Vinogradov distribution method \cite{vinogradov2009,Vinogradov2012} was applied, obtaining a total crosstalk probability of p = 0.37. This parameter will be used in Section~\ref{sec:scint} for the correction of the photoelectron yield.

\section{Scintillation light detection}
\label{sec:scint}

After the calibration described in Section~\ref{sec:calib}, the SiPM matrix and the \textit{cold section} of the electronics (Section~\ref{sec:elec}) were coupled to the crystal as described in Section~\ref{sec:crystal} and Figure~\ref{fig:detector}.
While the final goal of ASTAROTH is to have SiPM matrices on all six faces, the current work was performed with only one face instrumented. 
The other faces were wrapped in white PTFE tape serving as a diffuse reflector, thereby improving the overall light collection.

The detector was then installed in the cryostat (Section~\ref{sec:cryo}) and slowly cooled close to 80~K. Preliminary measurements were performed by increasing the temperature in steps of about 20~K.
In this paper the results based on data acquired at the lowest temperature are presented\footnote{A complete study of the temperature dependence of the system response, including laser calibrations performed by injecting light inside the cryostat, is planned and will be published separately.}.

To determine the energy scale and assess detector response, a \SI{47}{\kilo\becquerel} \ce{^{241}Am} radioactive source was placed \SI{24}{\cm} from one face of the detector. Two runs were acquired: one with the source and one for background only. The distance was chosen to yield a trigger rate of approximately 60~cps, nearly twice the background rate measured in the absence of the source.

The trigger threshold for each quadrant was set to be \qty{1.5}{\milli\volt}, approximately two-thirds of the single photoelectron pulse amplitude (see Figure~\ref{fig:persistency}), and comfortably above the typical RMS noise level of \qty{0.5}{\milli\volt}.
By means of the CAEN V1730 digitizer proprietary software, a 4-way coincidence among the quadrants within a \qty{48}{\nano\second} temporal window was implemented to isolate scintillation events from noise. The acquisition window for each event was set at \qty{12}{\micro\second}. 

A total of 40,000 waveforms were acquired per run and used to compute the charge deposited by scintillation events, obtained by subtracting the baseline and integrating over a \qty{5}{\micro\second} window starting at the trigger time. 
Assuming a photon emission time constant of approximately \qty{1.5}{\micro\second} for NaI(Tl) at cryogenic temperature~\cite{Sailer2012,Sibczynski2011}, the chosen integration\break window captures most of the scintillation signal. The charge measured in each quadrant was then converted to photoelectrons using the calibration described in Section~\ref{sec:calib_calib}. Finally, the total event charge in photoelectrons was obtained by summing the contributions from all four channels.
\subsection{Photoelectron yield evaluation}
Figure~\ref{fig:source} (a) shows the background spectrum (magenta) superimposed on the spectrum from the source run (blue). 
The main peak in the source spectrum is due to the 59.5~keV gamma of $^{241}$Am. A Gaussian fit to this peak returns a centroid at 430~photoelectrons (phe). Consequently, the gross photoelectron yield of this detector configuration is approximately 7.2~phe/keV. After correcting for the crosstalk probability, following the Vinogradov distribution as mentioned in Section~\ref{sec:calib}, the net yield was found to be approximately 4.5~phe/keV. The detector resolution, expressed as full width at half maximum (FWHM), was measured to be (26.6~$\pm$~0.4)~\% at 59.5~keV.
Based on the \ce{^241Am} energy calibration, the trigger condition of four photoelectrons corresponds to approximately 0.5~keV. It is worth emphasizing that the light yield calculation may be affected by several systematic uncertainties. In particular, the detector geometry plays a significant role, since the number of detected photons depends on the coupling between the SiPM and the crystal, as well as on the quality of the diffusive PTFE wrapping applied to the crystal. Moreover, each face of the crystal may exhibit slightly different optical properties, which can further influence the light yield. Therefore, the value reported in this measurement should be regarded as representative of the specific setup and operating conditions used. A calibration of the detector with a radioactive source is thus mandatory for each experimental run in which the system is reassembled.

The photoelectron yield was used to construct the upper x-axis of both panels in Figure~\ref{fig:source}. The secondary structure, extending approximately from 20~keV to 40~keV is attributed to the decay sub-products of \ce{^241Am}, such as \ce{^237Np}, which cannot be fully resolved with the current resolution.

The population of events below 10~keV is discussed in the following section. 

\begin{figure}
\resizebox{\columnwidth}{!}{%
\includegraphics{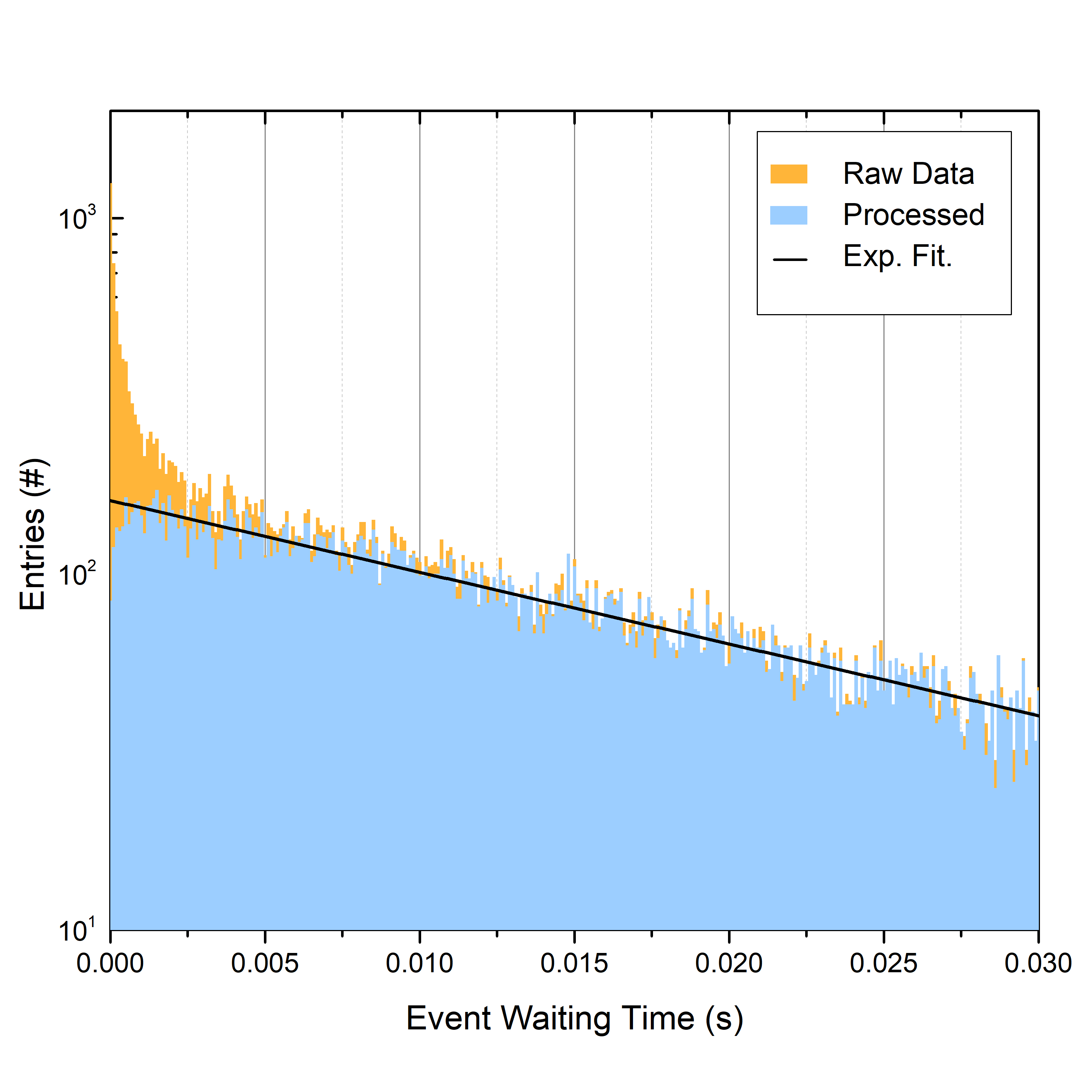}
}
\caption{Distribution of waiting times between events in both raw (yellow) and selected (blue) data. An exponential fit is performed on the selected data sample. See text for details.}
\label{fig:Dtime}
\end{figure}

\begin{figure*}
\resizebox{\textwidth}{!}{%
\includegraphics{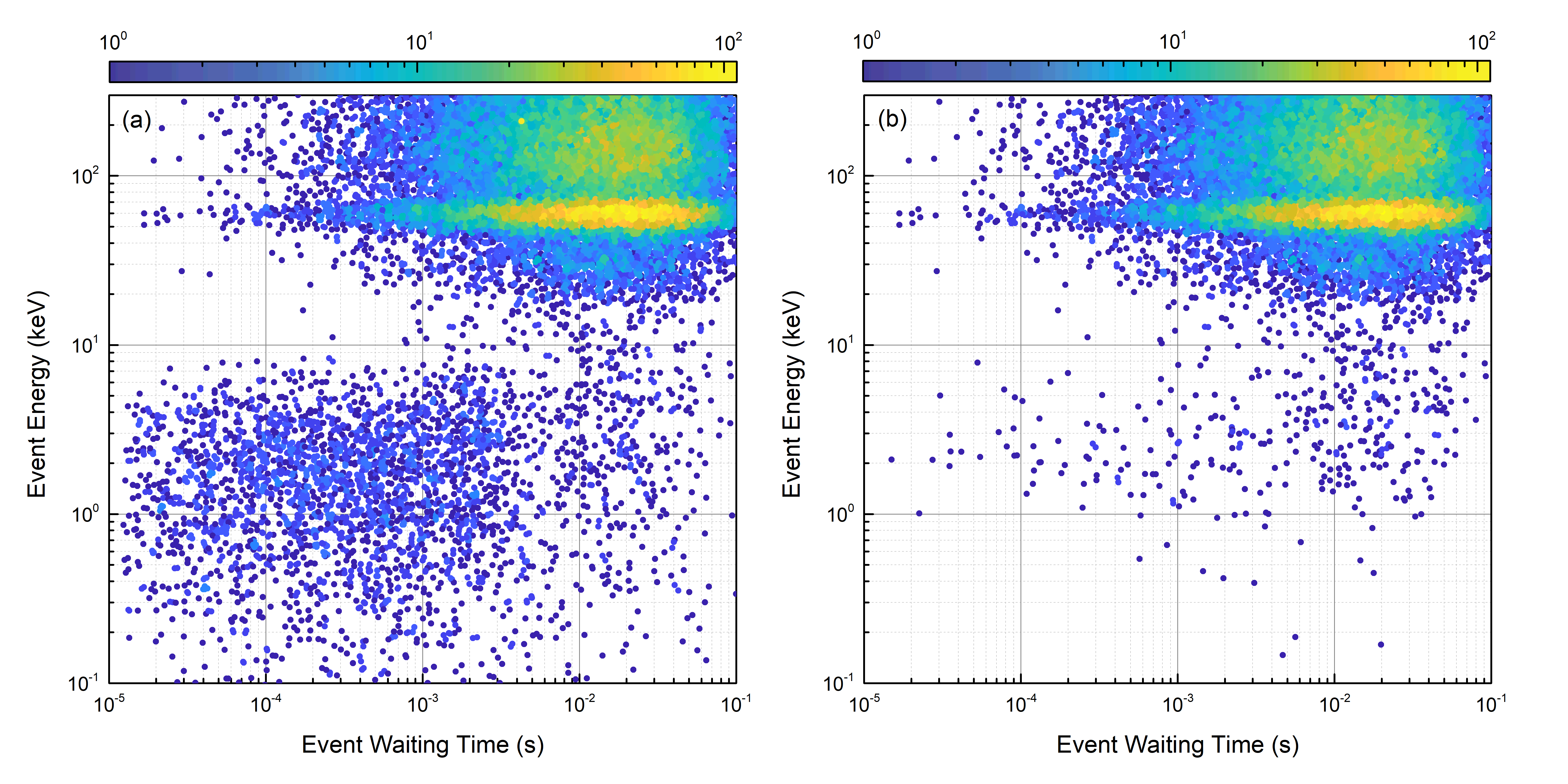}
}
\caption{Scatter plot correlating the waiting time between events with their energies for (a) all events and (b) events that pass the data quality selection.}

\label{fig:scatter}
\end{figure*}
 
\subsection{Data analysis}
\label{sec:filter}

As shown in Figure~\ref{fig:source} (b), the raw data exhibit a large number of events in the sub-10~keV energy region. To investigate this contribution, a histogram of waiting times — defined as the time of an event relative to the previous one — was constructed and is shown in Figure~\ref{fig:Dtime}. Poisson processes, such as those governing radioactive decay, are expected to show an exponential distribution of waiting times. Instead, the data show a clear excess above the expected exponential trend at short times.

The scatter plot shown in Figure~\ref{fig:scatter} (a) correlates the event energy with the waiting time. Short waiting times predominantly correspond to low-energy events.
In fact, such events are not true scintillation signals but rather artifacts caused by three primary factors.

The first occurs when one or more quadrants remain above threshold due to the lingering tail of a preceding pulse, falsely contributing to determine the trigger coincidence condition.
These events can be identified from the computation of the baseline in the \qty{2}{\micro\second} pre-trigger window (Section~\ref{sec:calib}): all those events whose baseline computation is anomalous due to the tail of the preceding event are discarded.
As a result of this procedure, 12.9~\% of the events were excluded from the source dataset and 6~\% for the background one.

The second occurrence is due to electronic noise events originating from a laboratory power device. This generates oscillatory bursts at around 20~MHz that propagate through the grounding system. To exclude those events, a moving-average low-pass filter with a cutoff frequency of approximately 10 MHz was applied to the data, and the same coincidence condition across all four quadrants was re-applied, resulting in the rejection of approximately 1~\% of the acquired waveforms.

The third occurrence is due to high-energy events, possibly induced by muons or other cosmic radiation. Following these events, the crystal emits a cascade of delayed photons extending over several milliseconds that repeatedly trigger the data acquisition and incorrectly appear as low-energy events. To prevent this contribution, all events within 500~ms after a high-energy (saturating) event were removed. Approximately 0.25~\% of the recorded waveforms are discarded this way and the dead time introduced is 0.2~\%.

As shown in Figure~\ref{fig:source} (b), these three data quality cuts clear most of the spurious population at low energies. The fraction of events discarded below 10~keV is 93.5~\%. On the other hand, above this energy where the vast majority of events are actual scintillation of the crystal, a mere 1.5~\% of events are removed by our data quality selection. The same reduction of the low energy spurious population can be observed in Figure~\ref{fig:scatter} (b).

As shown in Figure~\ref{fig:Dtime}, after the application of the data quality selection the distribution of waiting times between events can be fitted with an exponential curve. As further validation, the event rate returned by the fit is $\lambda_{\mathrm{fit}}$~=~46.4~cps, which compares well with the rate of events after the data selection, namely $\lambda_{\mathrm{sel}}$~=~48.4~cps. On contrary, the data acquisition rate before the quality selection was $\lambda_{\mathrm{raw}}$~=~56.6~cps. 

A $\chi^2$ test on the distributions of event counts occurring in each \SI{1}{\second} interval of the data acquisition was performed, for both the raw and filtered datasets. 
The raw data yielded a p-value of $10^{-8}$, indicating a significant deviation from the expected Poisson distribution. In contrast, the filtered data resulted in a p-value of 0.40, suggesting statistical compatibility.

As shown in Figure~\ref{fig:source}, a small fraction of low energy events survive the data quality selection for both the $^{241}$Am source and the background runs. To exclude that this signal is due to noise generated by the SiPMs, an additional run was performed where the SiPM matrix face was covered by black material while keeping everything else unaltered. The run lasted 17 hours and only a few triggers occurred, all of which were attributed to the grounding interference described above. This confirms the extremely low noise of SiPMs at cryogenic temperatures. Since there is no environmental light entering the cryostat, the residual low-energy events are attributed to light generated inside the setup, the origin of which requires further investigation. 

\section{Summary and Outlook}
\label{sec:conclusions}

ASTAROTH is an R\&D project aimed at developing a next-generation NaI(Tl) detector for direct dark matter searches.
The goal is to surpass PMT-based designs and achieve a substantial improvement in the signal-to-noise ratio at keV recoil energies by collecting scintillation light with large-area cryogenic SiPM matrices .

The target configuration employs cubic NaI(Tl) crystals with 5~cm sides, where scintillation light is read out by \break SiPMs on all faces. The detector operates at cryogenic temperatures (80–150~K) in a custom-designed cryostat. At these temperatures, SiPMs exhibit a strong reduction in dark noise compared to PMTs and offer advantages in terms of lower noise, higher photon detection efficiency, compactness, and reduced intrinsic radioactivity. The cryostat consists of a dual-walled copper chamber, immersed in a cryogenic bath and filled with helium gas, capable of hosting up to two detectors.

This paper presents the results of a measurement campaign conducted at the LASA laboratory with a minimal yet fully operational detector, consisting of a cylindrical NaI(Tl) crystal (5~cm diameter, 5~cm height) enclosed in a fused-silica cubic case. The crystal was read out on one face by a full-size SiPM matrix, while the remaining faces were\break wrapped in diffusive material.

Using a \ce{^{241}Am} cryogenic source, a net photoelectron yield of approximately \SI{4.5}{phe\per\kilo\electronvolt} was measured after crosstalk correction.
Noise in the low-energy region (below 10~keV) was effectively suppressed through data quality cuts.

These first results are encouraging and, to the best of our knowledge, represent the first published demonstration of the effectiveness of this technology. The next steps of the project include:
(1) transitioning to cubic crystals enclosed in custom epoxy resin cases;
(2) optically coupling SiPMs to the detector and extending the coverage to multiple faces;
(3) performing an independent study of the temperature dependence of the crystals scintillation light yield and pulse shape\footnote{On a longer time frame a measurement of the nuclear quenching factor at cold is also foreseen. To our knowledge this has not been measured so far.};
(4) adopting liquid argon as a coolant, which is also an efficient scintillator (40~photons/keV, peaked at 127~nm~\cite{Heindl_2010,Doke_2002}). By instrumenting the outer volume with additional SiPMs, it will operate as a veto detector to suppress key gamma-accompanied backgrounds, a technique so far implemented only with organic scintillators~\cite{cosine2021veto,SABREPoP}; and
(5) relocating the detector to an underground laboratory for a final and comprehensive characterization, shielded from cosmic radiation.

The ASTAROTH project proceeds in parallel with international efforts to test the long-standing DAMA claim of dark matter detection.
This work marks a key milestone, demonstrating that this technology holds strong potential for next-generation dark matter detectors.

\begin{acknowledgements}
    This work has been supported by the Fifth Scientific Commission (CSN5) of the Italian National Institute for Nuclear Physics (INFN), through the ASTAROTH grant.
    The authors thank the SABRE-North and DarkSide collaborations for fruitful discussions and exchange of know-how, as well as Fondazione Bruno Kessler for the co-development of the SiPM matrix. 
    They thank also Alessandro Andreani (Università degli Studi di Milano) for technical support, Sergio Brambilla (INFN, Milan) for assistance with data acquisition, Nicola Rossi (INFN, Laboratori Nazionali del Gran Sasso) for helping with the correction of crosstalk and Naomi Omori for reviewing the manuscript for style and language.
\end{acknowledgements}

\bibliographystyle{spphys}       
\bibliography{biblio}   

\begin{thebibliography}{10}
\providecommand{\url}[1]{{#1}}
\providecommand{\urlprefix}{URL }
\expandafter\ifx\csname urlstyle\endcsname\relax
  \providecommand{\doi}[1]{DOI \discretionary{}{}{}#1}\else
  \providecommand{\doi}{DOI \discretionary{}{}{}\begingroup \urlstyle{rm}\Url}\fi

\bibitem{Sofue2001}
Y.~Sofue, V.~Rubin, Ann. Rev. Astron. Astrophys. \textbf{39}, 137 (2001).
\newblock \urlprefix\url{https://doi.org/10.1146/annurev.astro.39.1.137}

\bibitem{Clowe2006}
D.~Clowe, et~al., Astrophys. J. \textbf{648}(2), L109 (2006).
\newblock \urlprefix\url{https://doi.org/10.1086/508162}

\bibitem{Freese2009}
{K. Freese}, EAS Publ. Ser. \textbf{36}, 113 (2009).
\newblock \urlprefix\url{https://doi.org/10.1051/eas/0936016}

\bibitem{Boylan2009}
M.~Boylan-Kolchin, V.~Springel, S.D.M. White, A.~Jenkins, G.~Lemson, Mon. Not. R. Astron. Soc. \textbf{398}(3), 1150 (2009).
\newblock \urlprefix\url{https://doi.org/10.1111/j.1365-2966.2009.15191.x}

\bibitem{Bertone2010}
{G. Bertone} (ed.), \emph{{Particle Dark Matter: Observations, Models and Searches}} (Cambridge University Press, 2010)

\bibitem{Goodman1985}
M.~W.Goodman, W.~Edward, Phys. Rev. D \textbf{31}, 3059 (1985).
\newblock \urlprefix\url{https://doi.org/10.1103/PhysRevD.31.3059}

\bibitem{cirelli2024}
M.~Cirelli, A.~Strumia, J.~Zupan.
\newblock Dark matter (2024).
\newblock \urlprefix\url{https://arxiv.org/abs/2406.01705}

\bibitem{Lewin1996}
{J.D. Lewin and P.~F. Smith}, Astropart. Phys. \textbf{6}(1), 87 (1996).
\newblock \urlprefix\url{https://doi.org/10.1016/S0927-6505(96)00047-3}

\bibitem{Freese2013}
K.~Freese, M.~Lisanti, C.~Savage, Rev. Mod. Phys. \textbf{85}, 1561 (2013).
\newblock \urlprefix\url{https://doi.org/10.1103/RevModPhys.85.1561}

\bibitem{Bernabei2013}
R.~Bernabei, et~al., Eur. Phys. J. C \textbf{73}, 1 (2013).
\newblock \urlprefix\url{https://doi.org/10.1140/epjc/s10052-013-2648-7}

\bibitem{Bernabei2018}
R.~Bernabei, et~al., Nucl. Phys. At. Energy \textbf{19}, 307 (2018).
\newblock \urlprefix\url{https://doi.org/10.15407/jnpae2018.04.307}

\bibitem{Bernabei2021}
R.~Bernabei, et~al., Nucl. Phys. At. Energy \textbf{22}, 329 (2021).
\newblock \urlprefix\url{https://doi.org/10.15407/jnpae2021.04.329}

\bibitem{Xenonnt}
E.~Aprile, et~al.
\newblock {WIMP} dark matter search using a 3.1 tonne $\times$ year exposure of the {XENONnT} experiment (2025).
\newblock \urlprefix\url{https://arxiv.org/abs/2502.18005}

\bibitem{PandaX2025}
B.~Zihao, et~al., Phys. Rev. Lett. \textbf{134}, 011805 (2025).
\newblock \urlprefix\url{https://doi.org/10.1103/PhysRevLett.134.011805}

\bibitem{LZ2024}
J.~Aalbers, et~al., Phys. Rev. Lett. \textbf{135}, 011802 (2025).
\newblock \urlprefix\url{https://doi.org/10.1103/4dyc-z8zf}

\bibitem{Agnes2018_2}
P.~Agnes, et~al., Phys. Rev. D \textbf{98}, 102006 (2018).
\newblock \urlprefix\url{https://doi.org/10.1103/PhysRevD.98.102006}

\bibitem{Catena2016}
R.~Catena, et~al., J. Cosm. Astropart. Phys. \textbf{2016}(05), 039 (2016).
\newblock \urlprefix\url{https://doi.org/10.1088/1475-7516/2016/05/039}

\bibitem{appec}
{Astroparticle Physics European Consortium}.
\newblock \url{https://www.appec.org/roadmap/}

\bibitem{anais2025}
J.~Amaré, et~al.
\newblock Towards a robust model-independent test of the {DAMA/LIBRA} dark matter signal: {ANAIS-112} results with six years of data (2025).
\newblock \urlprefix\url{https://arxiv.org/abs/2502.01542}

\bibitem{cosine2025}
G.H. Yu, et~al.
\newblock Limits on {WIMP} dark matter with {NaI(Tl)} crystals in three years of {COSINE-100} data (2025).
\newblock \urlprefix\url{https://arxiv.org/abs/2501.13665}

\bibitem{Antonello2019}
M.~Antonello, et~al., Eur. Phys. J. C \textbf{79}, 363 (2019).
\newblock \urlprefix\url{https://doi.org/10.1140/epjc/s10052-019-6860-y}

\bibitem{Calaprice2022}
F.~Calaprice, et~al., Eur. Phys. J. C \textbf{82} (2022).
\newblock \urlprefix\url{https://doi.org/10.1140/epjc/s10052-022-11108-z}

\bibitem{Fushimi2022}
K.~Fushimi, et~al., J. Phys. Conf. Ser. \textbf{2156} (2022).
\newblock \urlprefix\url{https://doi.org/10.1088/1742-6596/2156/1/012045}

\bibitem{Mariani21}
A.~Mariani, The proof-of-principle of the sabre experiment for the search of galactic dark matter through annual modulation.
\newblock Phd thesis, Gran Sasso Science Institute (2021).
\newblock \url{https://hdl.handle.net/20.500.12571/21981}

\bibitem{PandaXPMTS2024}
Y.~Yun, et~al., Nucl. Instrum. Methods Phys. Res. A \textbf{1073}, 170290 (2025).
\newblock \urlprefix\url{https://doi.org/10.1016/j.nima.2025.170290}

\bibitem{Dinu2016}
N.~Dinu, in \emph{Photodetectors}, ed. by B.~Nabet (Woodhead Publishing, 2016), pp. 255--294.
\newblock \urlprefix\url{https://doi.org/10.1016/B978-1-78242-445-1.00008-7}

\bibitem{Buzhan2003}
P.~Buzhan, et~al., Nucl. Instrum. Methods Phys. Res. A \textbf{504}, 48 (2003).
\newblock \urlprefix\url{https://doi.org/10.1016/S0168-9002(03)00749-6}

\bibitem{Biroth2015}
M.~Biroth, et~al., Nucl. Instrum. Methods Phys. Res. A \textbf{787}, 68 (2015).
\newblock \urlprefix\url{https://doi.org/10.1016/j.nima.2014.11.020}

\bibitem{Merzi_2023}
S.~Merzi, et~al., J. Instrum. \textbf{18}(05), P05040 (2023).
\newblock \urlprefix\url{https://dx.doi.org/10.1088/1748-0221/18/05/P05040}

\bibitem{Sibczynski2011}
P.~Sibczynski, et~al., in \emph{2011 IEEE Nuclear Science Symposium Conference Record} (IEEE, 2011), pp. 1616--1620.
\newblock \urlprefix\url{https://doi.org/10.1109/NSSMIC.2011.6154645}

\bibitem{SABRESouthPMTS2025}
O.~Stanley, et~al.
\newblock Characterisation of {Hamamatsu R11065-20 PMTs} for use in the {SABRE South} {NaI(Tl)} crystal detectors (2025).
\newblock \url{https://arxiv.org/abs/2504.17209}

\bibitem{Dangelo2023}
D.~D'Angelo, et~al., in \emph{AIP Conference Proceedings}, vol. 2908 (American Institute of Physics Inc., 2023), vol. 2908.
\newblock \urlprefix\url{https://doi.org/10.1063/5.0161723}

\bibitem{Anais_plus2024}
J.A. Allué, in \emph{15th International Workshop on the Identification of Dark Matter 2024 - IDM} (2024)

\bibitem{Knoll2010}
G.F. Knoll, \emph{{Radiation detection and measurement; 4th ed.}} (Wiley, New York, NY, 2010)

\bibitem{Sasaki2006}
S.~Sasaki, et~al., Jpn. J. Appl. Phys. \textbf{45}(8R), 6420 (2006).
\newblock \urlprefix\url{https://doi.org/10.1143/JJAP.45.6420}

\bibitem{Gallice23}
N.~Gallice, Cryogenic detection of scintillation light with large area {SiPM} arrays for next generation neutrino and dark matter experiments.
\newblock Phd thesis, {U}niversit\`a degli {S}tudi di {M}ilano (2023).
\newblock \url{https://hdl.handle.net/2434/956878}

\bibitem{Galli2022}
M.~Galli, {Development and characterisation of detectors based on ultra-high purity NaI(Tl) crystals for direct search of Dark Matter}.
\newblock Master's thesis, Universit\`a degli Studi di Milano (2022)

\bibitem{Acerbi2017}
F.~Acerbi, et~al., IEEE Trans. Electron Devices \textbf{64}, 521 (2017).
\newblock \urlprefix\url{https://doi.org/10.1109/TED.2016.2641586}

\bibitem{astaroth-simul}
F.~Alessandria, et~al.
\newblock A nonlinear multiphysics model for the design validation of the astaroth copper-steel cryogenic chamber (2025).
\newblock \url{http://arxiv.org/abs/2511.22529}

\bibitem{lakeshore}
{Lakeshore Model 336 Cryogenic Temperature Controller Datasheet}.
\newblock \urlprefix\url{https://www.lakeshore.com/products/categories/overview/temperature-products/cryogenic-temperature-controllers/model-336-cryogenic-temperature-controller}

\bibitem{Pan2005}
M.~Pan, Cryogenics \textbf{45}(6), 463 (2005).
\newblock \urlprefix\url{https://doi.org/10.1016/j.cryogenics.2005.03.006}

\bibitem{Acerbi2019}
F.~Acerbi, S.~Gundacker, Nucl. Instrum. Methods Phys. Res. A \textbf{926}, 16 (2019).
\newblock \urlprefix\url{https://doi.org/10.1016/j.nima.2018.11.118}

\bibitem{Villa2015}
F.~Villa, et~al., IEEE Trans. Nucl. Sci. \textbf{62}, 1950 (2015).
\newblock \urlprefix\url{https://doi.org/10.1109/TNS.2015.2477716}

\bibitem{Marano2014}
D.~Marano, et~al., IEEE Trans. Nucl. Sci. \textbf{61}, 23 (2014).
\newblock \urlprefix\url{https://doi.org/10.1109/TNS.2013.2283231}

\bibitem{lmh6626}
{LMH6626 - Single/ Dual Ultra Low Noise Wideband Operational Amplifier Datasheet, Texas Instruments}.
\newblock \urlprefix\url{https://www.ti.com/product/LMH6626}

\bibitem{lmh6624}
{Texas Instruments LMH6624 - Single/ Dual Ultra Low Noise Wideband Operational Amplifier Datasheet}.
\newblock \urlprefix\url{https://www.ti.com/product/LMH6624}

\bibitem{ths4541}
{Texas Instruments THS4541 - High-Speed Differential I/O Amplifier Datasheet}.
\newblock \urlprefix\url{https://www.ti.com/product/THS4541}

\bibitem{hamamatsu}
{Hamamatsu M10306-30 Laser - C10196 Controller Datasheet}.
\newblock \urlprefix\url{https://www.hamamatsu.com/eu/en/product/lasers/laser-related-products/laser-diode-heads/M10306series.html}

\bibitem{caenv1730}
{CAEN V1730 - 16/8 Channel 14 bit 500 MS/s Digitizer Datasheet}.
\newblock \urlprefix\url{https://www.caen.it/products/v1730}

\bibitem{caena4818}
{CAEN A4818 - USB 3.0 to CONET2 Adapter Datasheet}.
\newblock \urlprefix\url{https://www.caen.it/products/a4818}

\bibitem{vinogradov2009}
S.~Vinogradov, et~al., in \emph{2009 IEEE Nuclear Science Symposium Conference Record (NSS/MIC)} (2009), pp. 1496--1500.
\newblock \urlprefix\url{https://doi.org/10.1109/NSSMIC.2009.5402300}

\bibitem{Vinogradov2012}
S.~Vinogradov, in \emph{Nuclear Instruments and Methods in Physics Research, Section A: Accelerators, Spectrometers, Detectors and Associated Equipment}, vol. 695 (2012), vol. 695, pp. 247--251.
\newblock \urlprefix\url{https://doi.org/10.1016/j.nima.2011.11.086}

\bibitem{Sailer2012}
C.~Sailer, et~al., Eur. Phys. J. C \textbf{72}, 1 (2012).
\newblock \urlprefix\url{https://doi.org/10.1140/epjc/s10052-012-2061-7}

\bibitem{Heindl_2010}
T.~Heindl, et~al., Europhys. Lett. \textbf{91}(6), 62002 (2010).
\newblock \urlprefix\url{https://doi.org/10.1209/0295-5075/91/62002}

\bibitem{Doke_2002}
D.~Tadayoshi, et~al., Jpn. J. Appl. Phys. \textbf{41}(3R), 1538 (2002).
\newblock \urlprefix\url{https://doi.org/10.1143/JJAP.41.1538}

\bibitem{cosine2021veto}
G.~Adhikari, et~al., Nucl. Instrum. Methods Phys. Res. A \textbf{1006}, 165431 (2021).
\newblock \urlprefix\url{https://doi.org/10.1016/j.nima.2021.165431}

\bibitem{SABREPoP}
F.~Calaprice, et~al., Phys. Rev. D \textbf{104}, L021302 (2021).
\newblock \urlprefix\url{https://doi.org/10.1103/PhysRevD.104.L021302}

\end{thebibliography}

\end{document}